\documentclass[aps,prd,preprint,tightenlines,floatfix,amsmath,amssymb]{revtex4-1}
\usepackage[english]{babel}
\usepackage{hyphenat}
\usepackage{longtable}
\usepackage{graphicx}
\usepackage{bm}
\usepackage{lineno}
\usepackage{tabularx}
\usepackage{xspace}
\usepackage[usenames,dvipsnames,svgnames,table]{xcolor}

\newcommand{\epem}{\ensuremath{e^+e^-}\xspace}
\newcommand{\tablesize}{\fontsize{9pt}{9pt}\selectfont}

\begin{document}
\normalsize
\parskip=5pt plus 1pt minus 1pt
\title{\textbf{Amplitude Analysis of the $D^+ \rightarrow K^0_S \pi^+ \pi^0$ Dalitz Plot}}
\author{
\begin{small}
\begin{center}
\vspace{-0.4cm}
M.~Ablikim$^{1}$, M.~N.~Achasov$^{8,a}$, X.~C.~Ai$^{1}$, O.~Albayrak$^{4}$, M.~Albrecht$^{3}$, D.~J.~Ambrose$^{41}$, F.~F.~An$^{1}$, Q.~An$^{42}$, J.~Z.~Bai$^{1}$, R.~Baldini Ferroli$^{19A}$
% ...
\\
\vspace{0.4cm}
(BESIII Collaboration)\\
\vspace{0.2cm}
\textit{
$^{1}$ Institute of High Energy Physics, Beijing 100049, People's Republic of China\\
% ...
$^{3}$ Bochum Ruhr-University, D-44780 Bochum, Germany\\
$^{4}$ Carnegie Mellon University, Pittsburgh, Pennsylvania 15213, USA\\
% ...
$^{8}$ G.I. Budker Institute of Nuclear Physics SB RAS (BINP), Novosibirsk 630090, Russia\\
% ...
$^{19}$ (A)INFN Laboratori Nazionali di Frascati, I-00044, Frascati, Italy; (B)INFN and University of Perugia, I-06100, Perugia, Italy\\
% ...
$^{41}$ University of Rochester, Rochester, New York 14627, USA\\
$^{42}$ University of Science and Technology of China, Hefei 230026, People's Republic of China\\
% ...
$^{a}$ Also at the Novosibirsk State University, Novosibirsk, 630090, Russia\\
% ...
}\end{center}
\end{small}
}
\noaffiliation
\date{\today}

\begin{abstract}
  We perform an analysis of the $D^+\rightarrow K^0_S \pi^+ \pi^0$ Dalitz plot using a data set of 2.92~fb$^{-1}$ of \epem collisions at the $\psi(3770)$ mass accumulated by the BESIII Experiment, in which 166694 candidate events are selected with a background of 15.1\%. The Dalitz plot is found to be well-represented by a combination of six quasi-two-body decay channels ($K^0_S\rho^+$, $K^0_S\rho(1450)^+$, $\overline{K}^{*0}\pi^+$, $\overline{K}_0(1430)^0\pi^+$, $\overline{K}(1680)^0\pi^+$, $\overline{\kappa}^0\pi^+$) plus a small non-resonant component. Using the fit fractions from this analysis, partial branching ratios are updated with higher precision than previous measurements.
\end{abstract}

\maketitle

\section{Introduction}
A clear understanding of final-state interactions in exclusive weak decays is an important ingredient in our ability to predict decay rates and to model the dynamics of two-body decays of charmed mesons. Final-state interactions can cause significant changes in decay rates, and can cause shifts in the phases of decay amplitudes. Clear experimental measurements can help refine theoretical models of these phenomena.

Three-body decays provide a rich laboratory in which to study the interferences between intermediate-state resonances. They also provide a direct probe of final-state interactions in certain decays. When a particle decays into three or more daughters, such as the decay of $D\rightarrow P_1 P_2 P_3$, where $P_i$ ($i$=1,2,3) represents a pseudo-scalar particle, intermediate resonances dominate the decay rate. Amplitudes are typically obtained with a Dalitz plot analysis technique \cite{Dalitz}, which uses the minimum number of independent observable quantities, and any variation in the population over the Dalitz plot shows dynamical rather than kinematical effects. This provides the opportunity to experimentally measure both the amplitudes and phases of the intermediate decay channels, which in turn allows us to deduce their relative branching fractions. These phase differences can even allow details about very broad resonances to be extracted by observing their interference with other intermediate states.

A large contribution from a $K\pi$ $S$-wave intermediate state has been observed in earlier experiments including MARKIII \cite{MARK3}, NA14 \cite{NA14}, E691 \cite{E691}, E687 \cite{E687}, E791 \cite{E791a,E791b}, and CLEO-c \cite{CLEOc} in the $D^+ \rightarrow K^- \pi^+ \pi^+$ decay. Both E791 and CLEO-c interpreted their data with a Model-Independent Partial Wave Analysis (MIPWA) and found a phase shift at low $K\pi$ mass to confirm the $\overline{\kappa}\pi$ component. Complementary to $D^+ \rightarrow K^- \pi^+ \pi^+$, the $D^+\rightarrow K^0_S \pi^+\pi^0$ decay is also a golden channel to study the $K\pi$ $S$-wave in $D$ decays.

The previous Dalitz plot analysis of $D^+\rightarrow K^0_S \pi^+\pi^0$ by MARKIII \cite{MARK3} included only two intermediate decay channels, $K^0_S\rho$ and $\overline{K}^{*0}\pi^+$, and was based on a small data set. A much larger data sample of \epem collisions at $\sqrt{s} \approx 3.773$ GeV has been accumulated with the BESIII detector running at the Beijing Electron-Positron Collider (BEPCII). With much larger statistics, it is possible to measure relative branching fractions more precisely and to find more intermediate resonances. In this paper, we present an improved Dalitz plot analysis of the $D^+\rightarrow K^0_S \pi^+\pi^0$ decay.

\section{Event Selection}
\label{sec:data}

This analysis is based on a data sample of 2.92~fb$^{-1}$ \cite{lum}, which was collected at the peak of the $\psi(3770)$ resonance. BEPCII/BESIII \cite{BES3} is a major upgrade of the BESII experiment at the BEPC accelerator \cite{BEPC}. The design peak luminosity of the double-ring \epem collider, BEPCII \cite{BEPC2}, is $10^{33}$~cm$^{-2}$s$^{-1}$ at a beam current of 0.93~A. The BESIII detector with a geometrical acceptance of 93\% of $4\pi$ consists of the following main components: 1) a small-celled, helium-based main drift chamber (MDC) with 43 layers. The average single wire resolution is 135~$\mu$m, and the momentum resolution for 1~GeV/c charged particle in a 1~T magnetic field is 0.5\%. The chamber also provides a measurement of the specific energy loss $dE/dx$ for charged particles; 2) an electromagnetic calorimeter (EMC) made of 6240 CsI(Tl) crystals arranged in a cylindrical shape (barrel) plus two endcaps. For 1.0~GeV photons, the energy resolution is 2.5\% in the barrel and 5\% in the endcaps, and the position resolution is 6~mm in the barrel and 9~mm in the endcaps; 3) a Time-Of-Flight system (TOF) for particle identification composed of a barrel part made of two layers with 88 pieces of 5~cm thick, 2.4~m long plastic scintillators in each layer, and two endcaps with 96 fan-shaped, 5~cm thick, plastic scintillators in each endcap. The time resolution is 80~ps in the barrel, and 110~ps in the endcaps, corresponding to better than a 2$\sigma$ K/$\pi$ separation for momenta below about 1~GeV/c; 4) a muon chamber system (MUC) made of 1000~m$^2$ of Resistive Plate Chambers (RPC) arranged in 9 layers in the barrel and 8 layers in the endcaps and incorporated in the return iron of the superconducting magnet. The position resolution is about 2~cm.

At the $\psi(3770)$, $D$ mesons are produced in the reaction $\epem \rightarrow \psi(3770)\rightarrow D\overline{D}$. A single $D^+$ (or $D^-$) is first reconstructed by its daughters. This analysis uses the $D^+\rightarrow K^0_S \pi^+\pi^0$ decay and its charge conjugate channel. If one event contains both a $D^+$ and $D^-$ candidate, it will be treated as two events.

$K^0_S$ candidates are detected through the decay $K^0_S\rightarrow \pi^+\pi^-$. The pions from the $K^0_S$ are identified by requiring their $dE/dx$ be within $4\sigma$ of the pion hypothesis. In order to improve the signal-to-background ratio, the decay vertex of $\pi^+\pi^-$ pairs is required to be more than 2 standard deviations in the measurement of the decay length away from the interaction point, and their invariant mass is required to be within 20~MeV of the mass of the $K^0_S$. They are then kinematically constrained to the $K^0_S$ mass.

Charged $\pi$ candidates are required to satisfy $\left|\cos\theta\right|<0.93$, where $\theta$ is the polar angle with respect to the beam, to ensure reliable main drift chamber measurements. Only the tracks with points of closest approach to the beam line that are within 10~cm of the interaction point in the beam direction, and within 1~cm in the plane perpendicular to the beam, are selected. TOF and $dE/dx$ information are combined to form particle identification confidence levels for $\pi$ and $K$ hypotheses. Pions are identified by requiring the pion probability to be larger than that for a kaon.

$\pi^0$ candidates are detected through the decay $\pi^0\rightarrow \gamma\gamma$. Energy deposited in the nearby TOF counters is included in the photon energy measurement to improve the reconstruction efficiency and energy resolution \cite{TOF-EMC}. Photon candidates in the barrel region ($\left|\cos\theta\right|<0.8$, where $\theta$ is the polar angle of the shower) of the EMC must have at least 25~MeV total energy deposition; those in the endcap region ($0.84<\left|\cos\theta\right|<0.92$) must have at least 50~MeV total energy deposition. All neutral showers must lie in a window of EMC time measured by the rising edge of the signal in the pre-amplifier electronics to reduce the number of fake $\pi^0$ from random electronics noise and to improve their resolution. Of each $\gamma\gamma$ pair, at least one $\gamma$ is required to be in the barrel EMC, and the $\gamma\gamma$ mass is required to satisfy $0.115\text{~GeV}<m(\gamma\gamma)<0.150\text{~GeV}$. The pair is then kinematically constrained to the $\pi^0$ mass.

After $K^0_S$, $\pi^+$ and $\pi^0$ candidates are selected, $D^+$ candidates are constructed using the requirement $-73\text{~MeV}<\Delta E<41\text{~MeV}$, where $\Delta E=E_D-E_b$, $E_D$ is the sum of the $K^0_S$, $\pi^+$ and $\pi^0$ candidate energies, and $E_b$ is the beam energy, in the center mass system of \epem. For multiple $D^+$ candidates, the candidate with the smallest $\left|\Delta E\right|$ is chosen. We then perform a kinematic fit in which the invariant mass of the $D^+$ candidate is constrained to the $D^+$ mass, and its recoiling mass $m_{rec}=\sqrt{(\mathbf{p}_{e^+e^-}-\mathbf{p}_D)^2}$ is allowed to vary, where $\mathbf{p}_{e^+e^-}$ is the four-momentum of the \epem system and $\mathbf{p}_D$ is the four-momentum of the reconstructed $D^+$ candidate. This ensures that all $D^+$ decays have the same amount of phase space, regardless of whether the recoiling mass is in the signal or sideband region.

\begin{figure}
  \centering
  \includegraphics[width=1\linewidth]{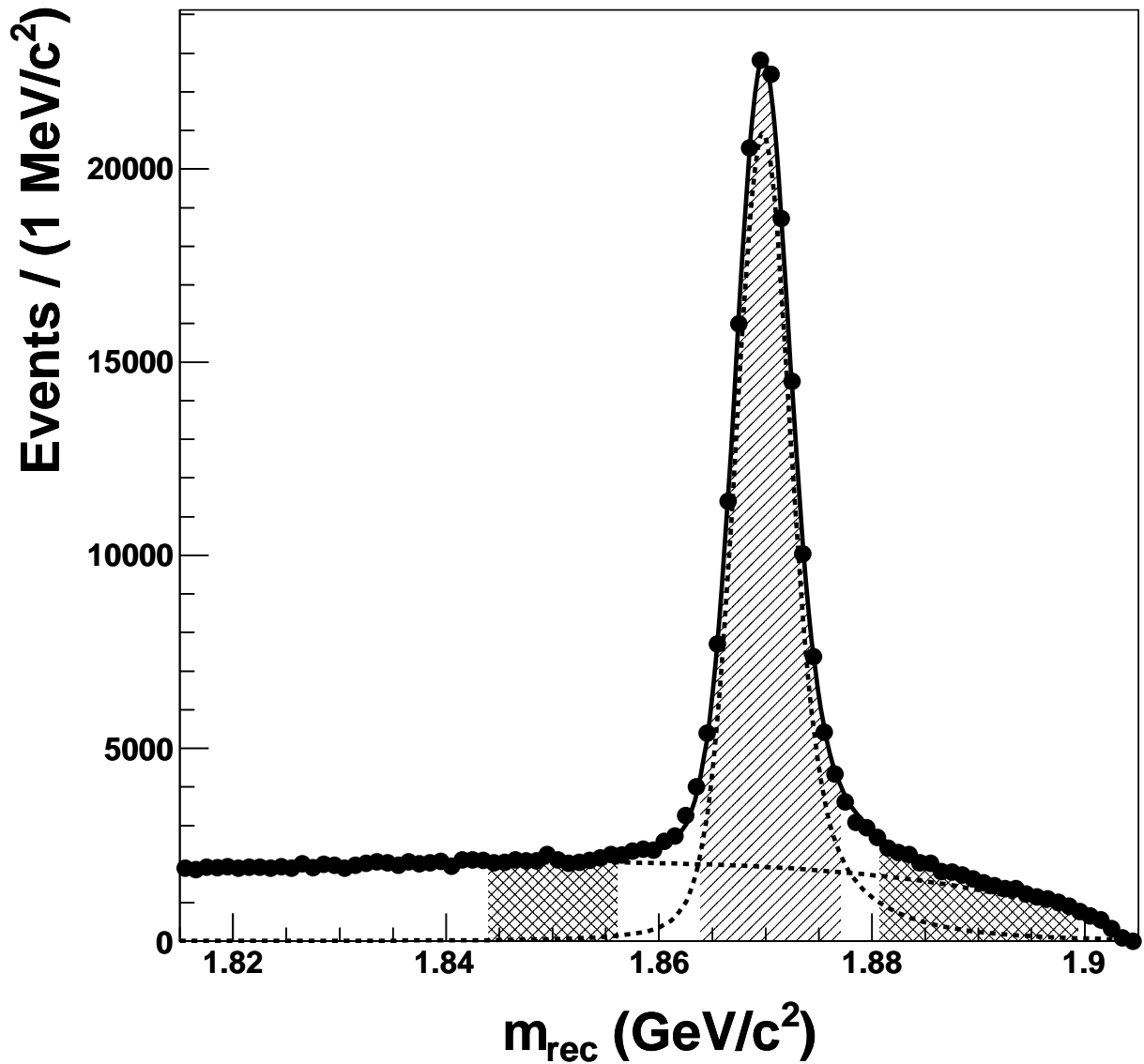}
  \caption{The recoil-mass distribution of $K^0_S\pi^+\pi^0$ candidates. \label{fig:mrec}}
\end{figure}

Figure~\ref{fig:mrec} shows the recoil-mass distribution fitted with a signal shape derived from Monte-Carlo (MC) simulation \cite{BES3mc}, with an ARGUS function~\cite{argus} for the combinatorial background. The signal shape is determined by the MC shape convolved with a Gaussian resolution function. The signal region is defined as $1.864\text{~GeV}<m_{rec}<1.877\text{~GeV}$, corresponding to the shaded region of Fig.~\ref{fig:mrec}; the events in the cross-hatched regions are taken as sideband events. In the signal region, the number of events above the combinatorial background is determined to be $142446\pm378$, and the amount of background in the signal region is estimated to be $24248\pm156$ events. Therefore, the size of the signal in the signal region is $(85.45\pm0.09)\%$ of the total. A peaking background contribution is included in the signal shape, which accounts for self-cross-feed events (where the $D^+$ decays to $K^0_S\pi^+\pi^0$, but the two $\pi^+$ are swapped, one from the $D^+$ and one from the $K^0_S$). The size of the peaking background is estimated using a MC study to be about $0.6\%$ of the signal size. Subtracting this background, the signal purity is $(84.9\pm0.1)\%$.

\begin{figure}
  \centering
  \includegraphics[width=1\linewidth]{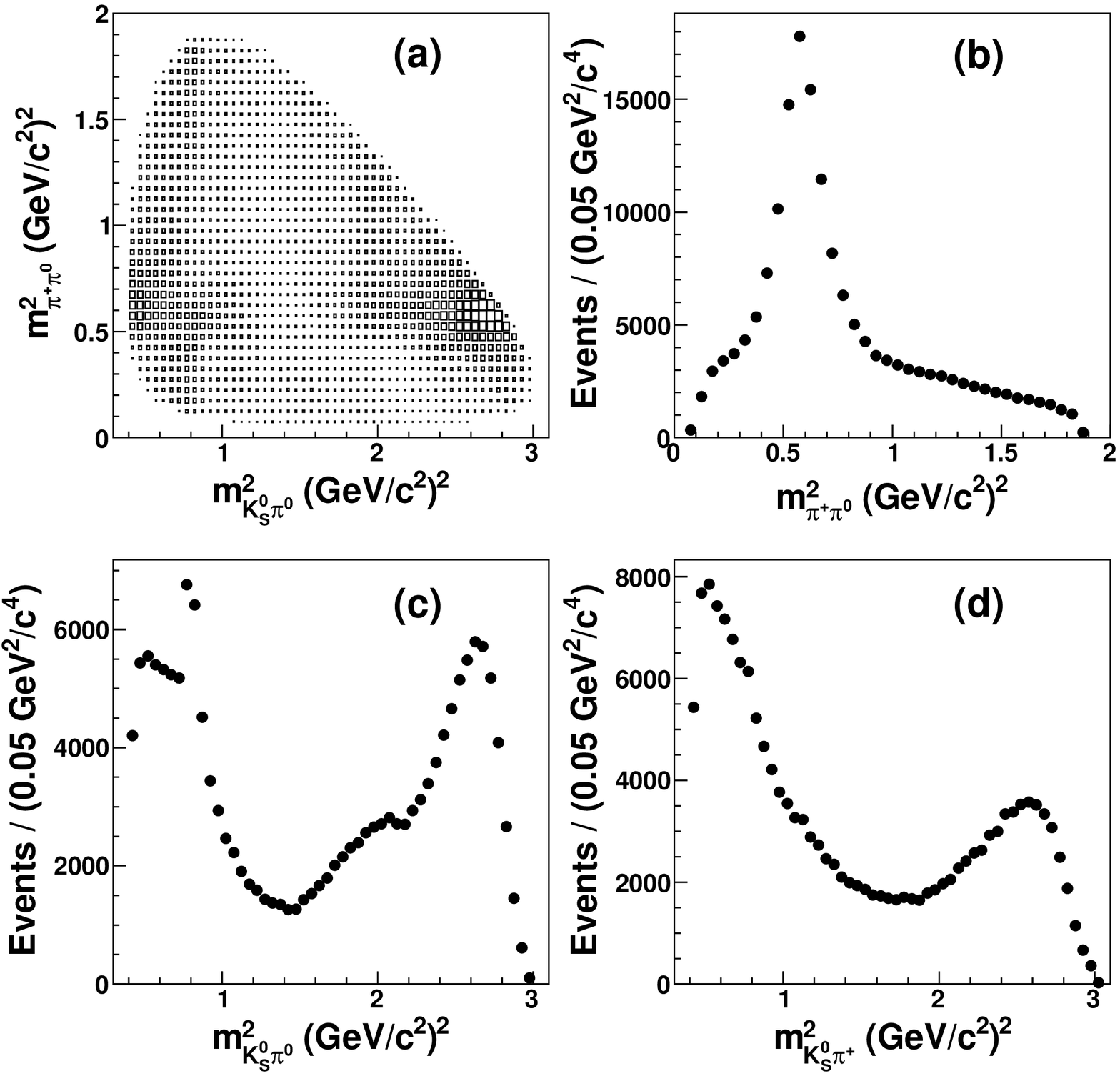}
  \caption{(a) The Dalitz plot for data and the projections onto (b) $m^2_{\pi^+\pi^0}$, (c) $m^2_{K^0_S\pi^0}$, and (d) $m^2_{K^0_S\pi^+}$.\label{fig:dalitz4}}
\end{figure}

Figure~\ref{fig:dalitz4} shows the Dalitz plot of the selected data sample in the signal region and three projections onto the squared $K^0_S\pi^0$ invariant mass ($m^2_{K^0_S\pi^0}$), the squared $\pi^+\pi^0$ invariant mass ($m^2_{\pi^+\pi^0}$), and the squared $K^0_S\pi^+$ invariant mass ($m^2_{K^0_S\pi^+}$). In this paper, $x=m^2_{K^0_S\pi^0}$ and $y=m^2_{\pi^+\pi^0}$ are selected as the two axes of the Dalitz plot, since only two of these three variables are independent according to energy and momentum conservation; $m^2_{K^0_S\pi^+}$ is defined as $z$.

\section{Partial Wave Analysis}
\subsection{Matrix element}

The $D^+\rightarrow K^0_S \pi^+ \pi^0$ Dalitz plot distribution satisfies $d\Gamma/dxdy\propto |\mathcal{M}|^2$, where $\mathcal{M}$ is the decay matrix element and contains the dynamics. The matrix element is parameterized by
\begin{equation}
\label{eqn:element}
\mathcal{M}=\sum^{L_{max}}_{L=0}Z_LF^L_DA_L,
\end{equation}
where $Z_L$ describes the angular distribution of the final-state particles; $F^L_D$ is the barrier factor for the production of the partial wave; and $A_L$ is the partial wave. The sum is over the decay orbital angular momentum $L$ of two-body partial waves. In this analysis we consider the sum up to the maximal orbital momentum $L_{max}=3$.

The partial waves $A_L$ are $L$-dependent functions of a single variable $s_R$ ($x$, $y$ or $z$). In the $D^+ \rightarrow K^0_S\pi^+\pi^0$ decay, the $S$, $P$, $D$ and $F$ waves ($L=0,1,2,3$
respectively) are represented by the sum of functions $\mathcal{W}_R$ for individual intermediate states:
\begin{subequations}
\label{eqn:waves}
\begin{eqnarray}
A_0(x)&=&c_{NR}+\mathcal{W}_{\overline{\kappa}^0}+\mathcal{W}_{\overline{K}^*_0(1430)^0} \nonumber \\
      &+&\left(\mathcal{W}_{\kappa^{+}}+\mathcal{W}_{K^{*}_{0}(1430)^{+}}\right)_{\text{DCS}},\\
A_1(x)&=&\mathcal{W}_{\overline{K}^{*0}}+\mathcal{W}_{\overline{K}^*(1410)^0}+\mathcal{W}_{\overline{K}^*(1680)^0} \nonumber \\
      &+&\left(\mathcal{W}_{K^{*+}}+\mathcal{W}_{K^{*}(1410)^{+}} +\mathcal{W}_{K^{*}(1680)^{+}}\right)_{\text{DCS}}, \text{~~~~~}\\
A_1(y)&=&\mathcal{W}_{\rho}+\mathcal{W}_{\rho(1450)}, \\
A_2(x)&=&\mathcal{W}_{\overline{K}^*_2(1430)^0}+\left(\mathcal{W}_{K^{*}_{2}(1430)^{+}}\right)_{\text{DCS}}, \text{ and} \\
A_3(x)&=&\mathcal{W}_{\overline{K}^*_3(1780)^0}+\left(\mathcal{W}_{K^{*}_{3}(1780)^{+}}\right)_{\text{DCS}},
\end{eqnarray}
\end{subequations}
where the subscripts denote the intermediate resonances (expressed by $R$ generally), and those in Doubly-Cabbibo Suppressed (DCS) channels are marked out. The contribution of non-resonant ($NR$) decays is represented by $c_{NR}=a_{NR}e^{i\phi_{NR}}$, a complex factor with two fit parameters for magnitude $a_{NR}$ and phase $\phi_{NR}$. For each resonance, the function
\begin{equation}
\label{eqn:res}
\mathcal{W}_R=c_R W_R F^L_R
\end{equation}
is the shape of an individual resonance, $W_R$, multiplied by the barrier factor in the resonance $R$ decay vertex, $F^L_R$, and the coupling factor, $c_R=a_R e^{i\phi_R}$.

In this analysis, the angular distribution $Z_L$, the barrier factor $F^L_D$ ($F^L_R$), and the resonance dynamical function $W_R$ are chosen as described in Appendix \ref{sec:isobar}.

\subsection{Maximum likelihood fit}
In order to describe the event density distribution on the Dalitz plot we use a probability density function (p.d.f.) $\mathcal{P}(x,y)$ described as follows:
\begin{widetext}
\fontsize{10pt}{10pt}\selectfont
\begin{equation}
\mathcal{P}(x,y)=
\begin{cases} \frac{\varepsilon(x,y)\left|\mathcal{M}(x,y)\right|^2}{\int\limits_{DP} \varepsilon(x,y)\left|\mathcal{M}(x,y)\right|^2dxdy} & \text{for efficiency,} \\
              \frac{B_1(x,y)}{\int\limits_{DP}B_1(x,y)dxdy}            & \text{for background, and}\\
              f_{S}\frac{|\mathcal{M}(x,y)|^2\varepsilon(x,y)}{\int\limits_{DP}|\mathcal{M}(x,y)|^2\varepsilon(x,y)dxdy} +f_{B1}\frac{B_1(x,y)}{\int\limits_{DP}B_1(x,y)dxdy} +f_{B2}\frac{B_2(x,y)}{\int\limits_{DP}B_2(x,y)dxdy} & \text{for signal with background,}
\end{cases}
\label{eqn:probability}
\end{equation}
\end{widetext}
where the $\varepsilon(x,y)$, $B_1(x,y)$, and $B_2(x,y)$ are functions representing the shapes of the efficiency, combinatorial background, and peaking background across the Dalitz plot, respectively; $f_{S}$, $f_{B1}$, and $f_{B2}$ are the fractions of signal, combinatorial background and peaking background under the constraint that $f_{S}+f_{B1}+f_{B2}\equiv 1$; and the integral limit $DP$ denotes the kinematic limit of the Dalitz plot. The p.d.f. free parameters are optimized with a maximum likelihood fit, where the log-likelihood function is described as
\begin{equation}
\ln \mathcal{L}=\sum^{N}_{i=1}\ln \mathcal{P}(x_i,y_i),
\label{eqn:likelihood}
\end{equation}
where $N$ is the number of events in the sample to parameterize.

Since we will test different models and obtain different parameters from different fits, we choose the Pearson goodness of fit to check them. A $\chi^2$ variable for the multinomial distribution on the binned Dalitz plot is defined as \begin{equation}
\chi^2=\sum^N_{i=1}\frac{(n_i-v_i)^2}{v_i},
\end{equation}
where $N$ is the number of the bins, $n_i$ is the number of events observed in the $i$th bin, and $v_i$ is the number predicted from the fitted p.d.f.

\subsection{Fit fractions}
We calculate the contribution of each component in the matrix element using a standard definition of the fit fraction
\begin{equation}
FF_C=\frac{\int\limits_{DP}\left|\sum\limits_{i\in C}\mathcal{A}_i(x,y)\right|^2dxdy}{\int\limits_{DP} |\mathcal{M}(x,y)|^2dxdy},
\label{eqn:ff}
\end{equation}
where $\mathcal{A}_i(x,y)$ is the amplitude contribution of the $i$th component, described as $c_R Z_L F^L_D F^L_R W_R$ for resonances and $c_{NR}$ for the non-resonant component, and $C$ is any combined set of components. When $C$ includes only one element, Eq. \ref{eqn:ff} gives the fit fraction of a single component. For the $K^0_S\pi^0$ $S$-wave, it consists of a non-resonant piece, a $K^*_0(1430)^0$, and a $\kappa^0$.

\subsection{Parameters}
In Eq. \ref{eqn:likelihood}, parameters include the ratios of signal and backgrounds, parameters describing the shapes of the efficiency and backgrounds, the coupling factors, the masses and widths of resonances, and the effective radii. Parameters for the efficiency shape are determined by studies of MC samples, and the backgrounds are estimated with the $m_{rec}$ sideband events of data. They are fixed in the fit to data. The ratios of signal and backgrounds are also fixed by first fitting the $m_{rec}$ distribution and studying the signal MC samples. The complex coupling factors, including the magnitude and phase, are free parameters in the fit and are used to calculate the fit fractions, but the magnitude and phase of the $K^0_S\rho^+$ component (which has the largest fit fraction) are fixed as 1 and 0. The masses and widths of the $\kappa$ and the $K^*_0(1430)$ are allowed to vary. Those of the other resonances used in the fit are fixed to their PDG \cite{PDG} values. The effective radii for the barrier factors are fixed at $r_D=5.0\text{~GeV}^{-1}$ and $r_R=1.5\text{~GeV}^{-1}$.

\section{Fitting Procedure}
\subsection{Efficiency}
We determine the efficiency for signal events as a function of position in the two-dimensional Dalitz plot, which can be described as a product of a polynomial function and threshold factors:
\begin{eqnarray}
\varepsilon(x,y)&=&T(v)(1+E_x x +E_y y +E_{xx} x^2+E_{xy}xy \nonumber \\
                &+&E_{yy} y^2+E_{xxx} x^3+E_{xxy} x^2 y +E_{xyy}xy^2  \nonumber \\
                &+&E_{yyy}y^3 ),
\end{eqnarray}
where $T(v)$ are the threshold factors for each Dalitz plot variable $v(x,y \text{ or } z)$, defined with an exponential form
\begin{equation}
T(v)=E_{0,v}+\left(1-E_{0,v}\right)\left[1-e^{-E_{th,v}\left|v-v_{edge}\right|}\right].
\end{equation}
All polynomial coefficients $E_x$, $E_y$, $E_{xx}$, $E_{xy}$, $E_{yy}$, $E_{xxx}$, $E_{xxy}$, $E_{xyy}$, and $E_{yyy}$ are fit parameters. In the threshold function, the parameter $E_{th,v}$ is free in the fit and $v_{edge}$ is defined as the expected value of $v$ at the Dalitz plot edge. $E_0$ denotes the efficiency when $v=v_{max}$. The threshold factor describes the low efficiency in regions with $v \rightarrow v_{max}$, where one of the three particles is produced with zero momentum in the $D$ meson rest frame. We consider the threshold for $v=m_{K^0_S\pi^0}^2$ and $v=m^2_{\pi^+\pi^0}$.

To determine the efficiency we use a signal MC simulation \cite{BES3mc} in which one of the charged $D$ mesons decays in the signal mode, while the other $D$ meson decays in all its known decay modes with proper branching fractions. These events are input into the BESIII detector simulation and are processed with the regular reconstruction package. The MC-generated events are required to pass the same selection requirements as data in the signal region, as shown in Fig.~\ref{fig:mrec}. A track-matching technology is applied to the MC events to select only the signal mode side and to avoid contamination from the other $D$ meson. Then the efficiency is obtained by fitting Eq. \ref{eqn:probability} to this sample with fixed $\mathcal{M}(x,y)$.

\subsection{Background}
\label{background}
\begin{figure}[b]
  \centering
  \includegraphics[width=1\linewidth]{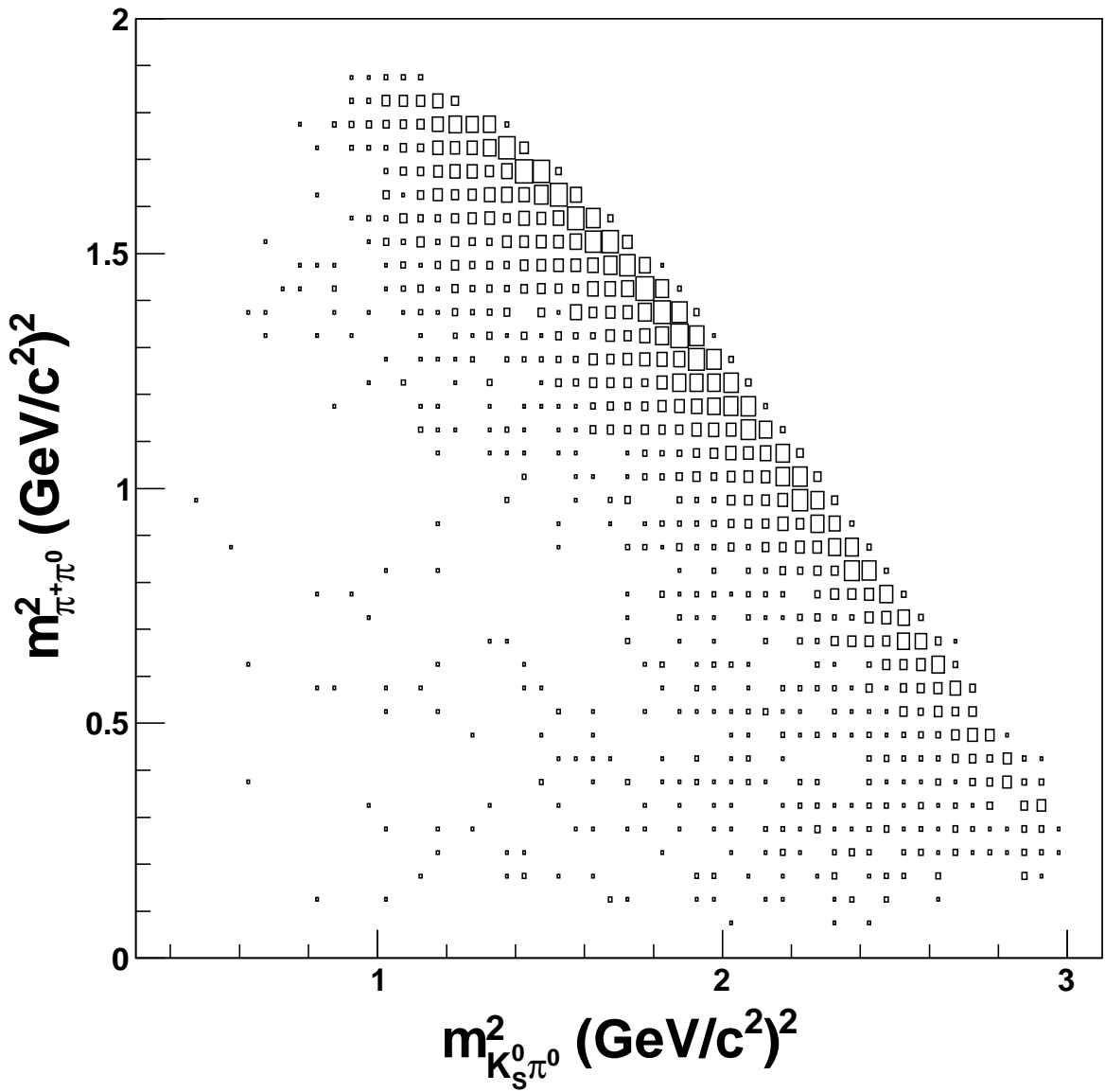}
  \caption{The shape of self-cross-feed events on the Dalitz plot.}
  \label{fig:mcbkg}
\end{figure}

As described in Section \ref{sec:data}, there are both combinatorial and peaking backgrounds. For the peaking background, the shape in the Dalitz plot is estimated by an MC sample, as shown in Fig.~\ref{fig:mcbkg}. Most of the self-cross-feed events have small $m^2_{K^0_S\pi^+}$ values, corresponding to small angles between the $K^0_S$ and the $\pi^+$. For the self-cross-feed contribution to the background, we use the histogram as the p.d.f.\ of $B_2(x,y)$. For the combinatorial background, we use data events from the two $m_{rec}$ sideband regions, shown by the hatched range in Fig.~\ref{fig:mrec}.

\begin{figure*}
  \centering
  \includegraphics[width=1\linewidth]{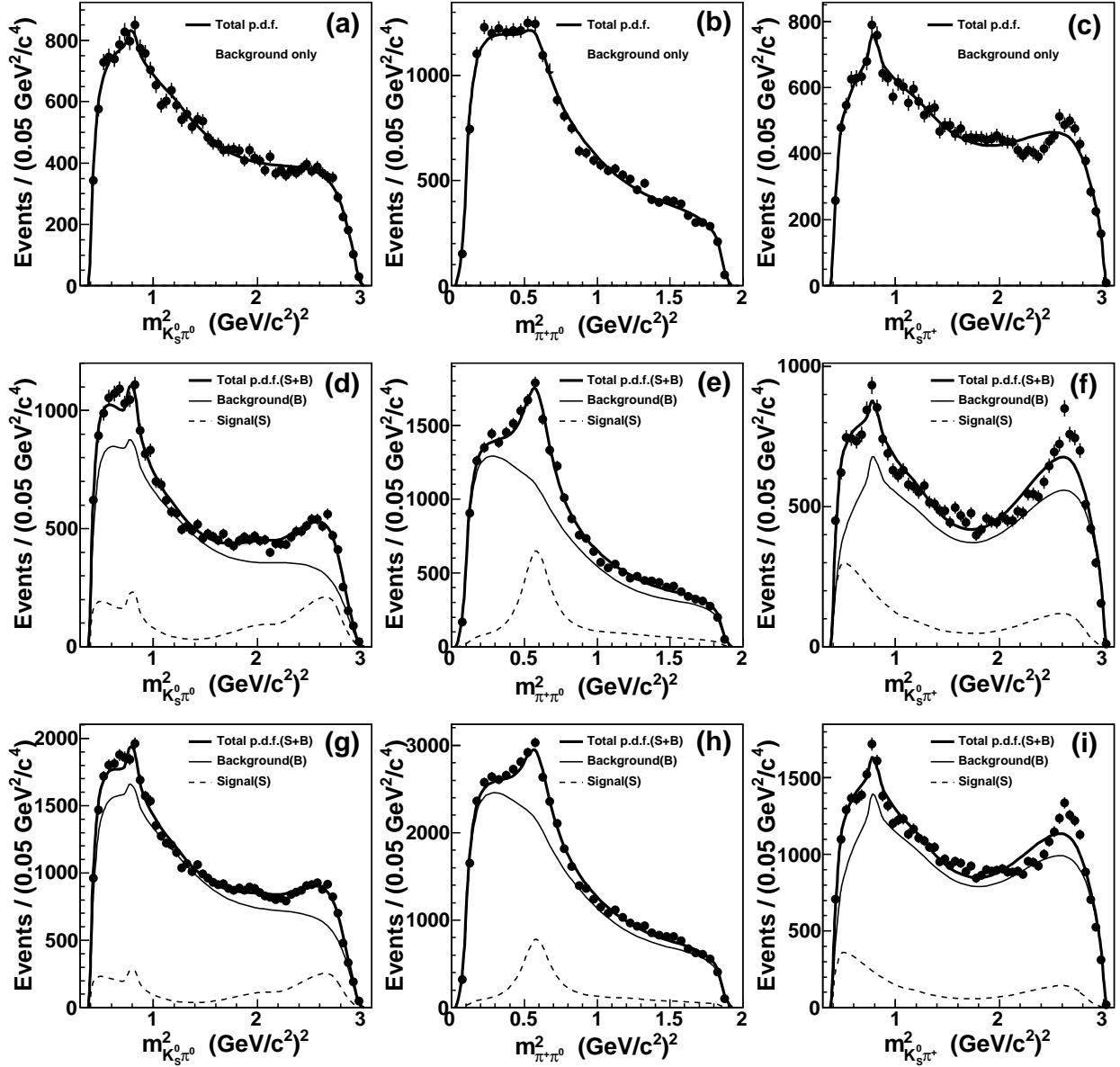}
  \caption{Results of the fit to the sideband backgrounds: (a), (b), and (c) are the three projections for the low-mass sideband only; (d), (e), and (f) are for the high-mass sideband only; and (g), (h), and (i) are for the combined sidebands. The signals are fed in for the signal tail in the sideband of the $m_{rec}$ distribution.}
  \label{fig:bkgfit}
\end{figure*}

Because the high-mass $m_{rec}$ sideband has a significant contribution from signal events due to a tail caused by initial state radiation, we consider a contribution of signal for these events, whose fraction is obtained by fitting the distribution of $m_{rec}$. The contribution of signal, $\mathcal{M}_0$, is initialized by the parameterized shape of the low-mass sideband, $\mathcal{B}_0$. The $\mathcal{B}_0$ is fitted by Eq. \ref{eqn:probability} for background using events in the low-mass sideband, and then the $\mathcal{M}_0$ is fitted using $\mathcal{B}_0$ as $B_1(x,y)$. After that, the events in both sidebands are used to estimate the shape of the background in the Dalitz plot. $\mathcal{B}_1$ is parameterized to the total background events in sidebands by Eq. \ref{eqn:probability} for signal with background, based on the fixed $\mathcal{M}_0$, and then the $\mathcal{M}_1$ is fitted using $\mathcal{B}_1$. In order to make sure the right resonance contribution is used, this process is repeated $i$ times to obtain $\mathcal{B}_i$ and $\mathcal{M}_i$ until the variation of the signal from the last result is small enough. In this analysis, this process is repeated once.

The dominant misreconstructed $D$ decays are from $D^+\rightarrow K^0_Sa_1(1260)^+$, $D^0\rightarrow K^-\pi^+\pi^0$, and $D^0\rightarrow K^0_S \pi^+\pi^-\pi^0$. It is worth noting that the background from the $D^0$ decay will bring a $K^*(892)^+$ contribution to the Dalitz plot which is a DCS process in the $D^+\rightarrow K^0_S\pi^+\pi^0$ decay.  We take this into account by adding this noncoherent $K^*(892)^+$ contribution to the background p.d.f., along with the $\rho(770)^+$ and $K^*(892)^0$ described below.

To parameterize the background shape on the Dalitz plot we employ a function similar to that used for the efficiency:
\begin{eqnarray}
B_1(x,y)&=&T(x)(1+B_x x +B_y y +B_{xx} x^2+B_{xy} xy \nonumber \\
        &+&B_{yy} y^2+B_{xxx} x^3 +B_{xxy}x^2y+B_{xyy}xy^2      \nonumber \\
        &+&B_{yyy}y^3+B_{\rho}|\mathcal{A}_{\rho}|^2+B_{\overline{K}^{*0}}\left|\mathcal{A}_{\overline{K}^{*0}}\right|^2 \nonumber \\
        &+&B_{\overline{K}^{*+}}\left|\mathcal{A}_{\overline{K}^{*+}}\right|^2),
\end{eqnarray}
where all the coefficients, $B_x$, $B_y$, $B_{xx}$, $B_{xy}$, $B_{yy}$, $B_{xxx}$, $B_{xxy}$, $B_{xyy}$, $B_{yyy}$, $B_{\rho}$, $B_{\overline{K}^{*0}}$, and $B_{\overline{K}^{*+}}$, are fit parameters. Unlike the efficiency parameterization, the terms for the intermediate resonances $\rho$ and $\overline{K}^{*0}$ describe the contributions from these resonances. Figure \ref{fig:bkgfit} shows the results of the fit with the background-corrected polynomial function to our sideband sample. There are some deviations between the parameterized functions and the sidebands, which primarily lie on the projection of $m^2_{K^0_S\pi^+}$. The deviations will be considered as one source of systematic error in Section \ref{bkgerr}. The impact of the deviation is comparable to other sources of systematic uncertainties.

\subsection{Fit to data}

\begin{table}[b]
  \centering
  \caption{The intermediate resonance decay modes considered in this analysis.}\label{tab:resonance}
  \begin{tabularx}{\linewidth}{|X|X|X|}
    \hline
    \hline
    \multicolumn{2}{|l|}{CF mode} & DCS mode\\
    \cline{1-3}
    $K^0_S X^+$          & $X^0\pi^+$ & $X^+\pi^0$ \\
    \hline
    \hline
    $K^0_S \rho(770)^+$  & $\overline{K}^*(892)^0\pi^+$    & $K^*(892)^+\pi^0$ \\
    $K^0_S \rho(1450)^+$ & $\overline{K}^*_0(1430)^0\pi^+$ & $K^*_0(1430)^+\pi^0$ \\
                         & $\overline{K}^*(1680)^0\pi^+$   & $K^*(1680)^+\pi^0$ \\
                         & $\overline{\kappa}^0\pi^+$      & $\kappa^{+}\pi^{0}$ \\
    \cline{1-3}
    $K^0_S \rho(1700)^+$ & $\overline{K}^*(1410)^0\pi^+$   & $K^*(1410)^+\pi^0$ \\
                         & $\overline{K}^*_2(1430)^0\pi^+$ & $K^*_2(1430)^+\pi^0$ \\
                         & $\overline{K}^*_3(1780)^0\pi^+$ &  $K^*_3(1780)^+\pi^0$ \\
    \hline
    \hline
  \end{tabularx}
\end{table}

A previous analysis from the MARKIII experiment \cite{MARK3} included only two intermediate resonances in the $D^+\rightarrow K^0_S\pi^+\pi^0$ decay: $K^0_S\rho^+$ and $\overline{K}^{*0}\pi^+$. Obvious contributions from more resonances have been seen in the more recent $D^+\rightarrow K^-\pi^+\pi^+$ analyses. Hence, more resonances are considered in this analysis. All possible intermediate resonance decay modes are listed in Table \ref{tab:resonance}, including Cabbibo Favored (CF) modes and DCS modes. A model using only these CF channels is found to be adequate. No evidence is found for additional DCS channels. However, the heavy $\rho$ mesons, $\rho(1450)$ and $\rho(1700)$, contribute parts of their resonance shapes, and then their shapes in the Dalitz plot are close. As pointed out by CLEO \cite{CLEO2}, the inclusion of both $\rho$ resonances is probably a misrepresentation of the contents of the Dalitz plot. In order to avoid fake interference, we choose only one of them, the $\rho(1450)$, to express approximatively their combined contribution in the decay matrix element. The results of the CF model (called model A) with a complex pole for the $\kappa$ and Breit-Wigner functions for the other resonances are listed in the column \textquotedblleft Model A\textquotedblright\,of Table \ref{tab:fitresult}.

\begin{table*}
  \centering
  \tablesize
  \caption{The results of the fits to the $D^+ \rightarrow K^0_S\pi^+\pi^0$ Dalitz plot with a complex pole for the $\kappa$ and Breit-Wigner functions for others, described in the text. The first term of errors are statistical and the second terms are experimental errors in Model A, and statistical only in Model B, C, and D. Model A includes all decay modes listed in the first column. Based on the Model A, Model B excludes the contribution of $\overline{\kappa}^0\pi^+$; Model C excludes the non-resonant contribution; Model D consists of the decay modes after dropping the modes with small fractions, $\overline{K}^*(1410)^0\pi^+$, $\overline{K}^*_2(1430)^0\pi^+$, and $\overline{K}^*_3(1780)^0\pi^+$. The $S$-wave is calculated by adding the non-resonant component, the $\overline{\kappa}^0\pi^+$, and the $\overline{K}^*_0(1430)^0\pi^+$.} \label{tab:fitresult}
  \begin{tabularx}{\linewidth}{|X|X|X|X|X|X|}
    \hline
    \hline
    Decay Mode & Par. & Model A & Model B & Model C & Model D \\
    \hline
    Non-resonant                    & FF(\%) & 4.5$\pm$0.7$\pm$2.6     & 18.3$\pm$0.6    &                 & 6.1$\pm$0.9   \\
                                    & $\phi$(\ensuremath{^{\circ}}) & 269$\pm$6$\pm$26       & 232.7$\pm$1.3   &                 & 276$\pm$6     \\
    $K^0_S\rho(770)^+$              & FF(\%) & 84.6$\pm$1.8$\pm$2.5    & 82.0$\pm$1.3    & 86.7$\pm$1.1    & 82.2$\pm$2.2  \\
                                    & $\phi$(\ensuremath{^{\circ}}) & 0(fixed)        & 0(fixed)        & 0(fixed)        & 0(fixed)      \\
    $K^0_S\rho(1450)^+$             & FF(\%) & 1.8$\pm$0.2$\pm$0.8   & 6.03$\pm$0.29   & 0.63$\pm$0.12   & 2.65$\pm$0.28 \\
                                    & $\phi$(\ensuremath{^{\circ}}) & 198$\pm$4$\pm$10       & 167.1$\pm$2.1   & 186$\pm$8       & 183.7$\pm$2.6 \\
    $\overline{K}^*(892)^0\pi^+$    & FF(\%) & 3.22$\pm$0.14$\pm$0.15   & 2.99$\pm$0.10   & 3.30$\pm$0.10   & 3.38$\pm$0.16 \\
                                    & $\phi$(\ensuremath{^{\circ}}) & 294.7$\pm$1.3$\pm$1.4   & 279.3$\pm$1.2   & 292.3$\pm$1.5   & 292.2$\pm$1.3 \\
    $\overline{K}^*(1410)^0\pi^+$   & FF(\%) & 0.12$\pm$0.05$\pm$0.17   & 0.18$\pm$0.05   & 0.12$\pm$0.05   &               \\
                                    & $\phi$(\ensuremath{^{\circ}}) & 228$\pm$9$\pm$26       & 301$\pm$10      & 243$\pm$12      &               \\
    $\overline{K}^*_0(1430)^0\pi^+$ & FF(\%) & 4.5$\pm$0.6$\pm$1.2     & 10.5$\pm$1.3    & 3.6$\pm$0.5     & 3.7$\pm$0.6   \\
                                    & $\phi$(\ensuremath{^{\circ}}) & 319$\pm$5$\pm$14       & 306.2$\pm$2.0   & 317$\pm$4       & 339$\pm$5     \\
                                    & mass(MeV) & 1452$\pm$5$\pm$15   & 1435$\pm$4      & 1449$\pm$4      & 1470$\pm$6    \\
                                    & width(MeV)& 184$\pm$7$\pm$15    & 287$\pm$11      & 163$\pm$6       & 187$\pm$7     \\
    $\overline{K}^*_2(1430)^0\pi^+$ & FF(\%) & 0.12$\pm$0.02$\pm$0.09 & 0.086$\pm$0.014 & 0.111$\pm$0.015 &               \\
                                    & $\phi$(\ensuremath{^{\circ}}) & 273$\pm$7$\pm$18       & 265$\pm$9       & 267$\pm$7       &               \\
    $\overline{K}^*(1680)^0\pi^+$   & FF(\%) & 0.21$\pm$0.06$\pm$0.08   & 0.58$\pm$0.08   & 0.43$\pm$0.10   & 1.05$\pm$0.09 \\
                                    & $\phi$(\ensuremath{^{\circ}}) & 243$\pm$6$\pm$22       & 284$\pm$4       & 234$\pm$5       & 255.3$\pm$2.0 \\
    $\overline{K}^*_3(1780)^0\pi^+$ & FF(\%) & 0.034$\pm$0.008$\pm$0.020 & 0.055$\pm$0.008 & 0.037$\pm$0.008 &               \\
                                    & $\phi$(\ensuremath{^{\circ}}) & 130$\pm$12$\pm$50      & 113$\pm$9       & 131$\pm$11      &               \\
    $\overline{\kappa}^0\pi^+$      & FF(\%) & 6.8$\pm$0.7$\pm$2.2     &                 & 18.8$\pm$0.5    & 6.4$\pm$1.0   \\
                                    & $\phi$(\ensuremath{^{\circ}}) & 92$\pm$6$\pm$22        &                 & 11.6$\pm$1.9    & 92$\pm$7      \\
                                    & $\Re$(MeV)& 739$\pm$14$\pm$40  &          & 773$\pm$11      & 750$\pm$15    \\
                                    & $\Im$(MeV)& -220$\pm$14$\pm$15 &          & -396$\pm$18     & -230$\pm$21   \\
    \hline
    $NR$+$\overline{\kappa}^0\pi^+$   & FF(\%) & 18.1$\pm$1.4$\pm$1.6    & 18.3$\pm$0.6    & 18.8$\pm$0.5    & 19.2$\pm$1.8  \\
    $K^0_S\pi^0$ $S$-wave           & FF(\%) & 18.9$\pm$1.0$\pm$2.0    & 15.8$\pm$1.0    & 21.2$\pm$1.0    & 17.1$\pm$1.4  \\
    \hline
     \multicolumn{2}{|c|}{$\Sigma$FF(\%)}  & 106       & 121       & 114       & 105       \\
     \multicolumn{2}{|c|}{$\chi^2/Ndof$}      & 1672/1187 & 2497/1191 & 1777/1189 & 2068/1193 \\
     \multicolumn{2}{|c|}{$-2\ln \mathcal{L}$} & 239415    & 240284    & 239521    & 239807    \\
    \hline
    \hline
  \end{tabularx}
\end{table*}

Based on the model A, we perform a fit with a model without the $\overline{\kappa}$ (called model B) as a test, as listed in the column \textquotedblleft Model B\textquotedblright\,of Table \ref{tab:fitresult}. It is found that the goodness of fit is worse than in the model A, which demonstrates the presence of $\overline{\kappa}$ in our data at high confidence level.

Similarly, we also test the model without the non-resonant component (called model C), and the results are listed in the column \textquotedblleft Model C\textquotedblright\,of Table \ref{tab:fitresult}. The resulting $\chi^2$ increases by 105 units over the model A, indicating that a non-resonant component is indeed present in our data.

\begin{figure*}
  \centering
  \includegraphics[width=0.75\linewidth]{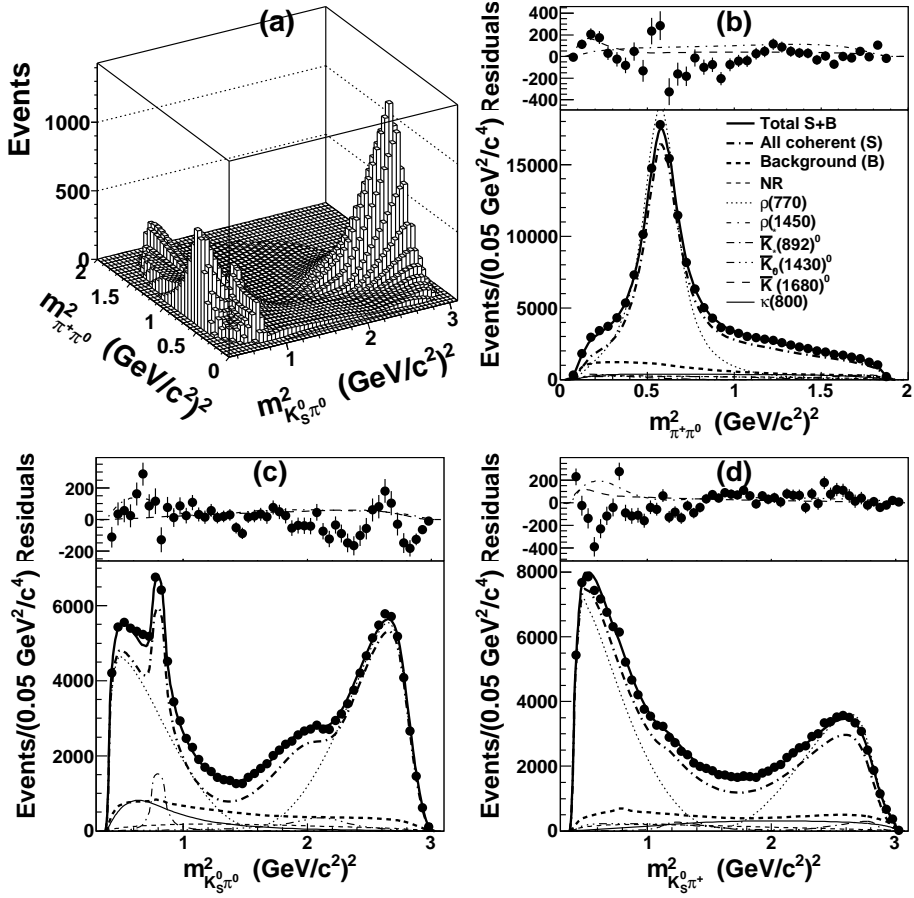}
  \caption{The results of fitting the $D^+\rightarrow K^0_S \pi^+\pi^0$ data with the model D. (a) Distribution of fitted p.d.f. and projections on (b) $m^2_{\pi^+\pi^0}$, (c) $m^2_{K^0_S\pi^0}$, and (d) $m^2_{K^0_S\pi^+}$. Residuals between the data and the total p.d.f. are shown by dots with statistical error bars in the top insets with minor contributions from the $\rho(1450)$ and the $\overline{K}^*(1680)^0$.}
  \label{fig:final}
\end{figure*}

In the above three models, the contributions of the three channels $\overline{K}^*(1410)^0\pi^+$, $\overline{K}^*_2(1430)^0\pi^+$ and $\overline{K}^*_3(1780)^0\pi^+$ are not significant, compared to the systematic uncertainties estimated in model A (listed in Table \ref{tab:fitresult}). Therefore, we remove them from the model A as the final model (called model D). The model D is composed of a non-resonant component and intermediate resonances, including $K^0_S\rho(770)^+$, $K^0_S\rho(1450)^+$, $\overline{K}^*(892)^0\pi^+$, $\overline{K}^*_0(1430)^0\pi^+$, $\overline{K}^*(1680)^0\pi^+$, and $\overline{\kappa}^0\pi^+$. The results are listed in the column \textquotedblleft Model D\textquotedblright\,of Table \ref{tab:fitresult}. Except for the large ($\sim$85\%) contributions from $K^{0}_{S}\rho(770)^{+}$ and $K^{0}_{S}\rho(1450)^{+}$, and a visible ($\sim$3\%) component of $\overline{K}^{*0}\pi^{+}$, a significant ($\sim$20\%) contribution of $K^{0}_{S}\pi^{0}$ $S$-wave is found in our fit.
The projections of the fit and the Dalitz plot is shown in Fig.~\ref{fig:final}.

\begin{figure}
  \centering
  \includegraphics[width=1\linewidth]{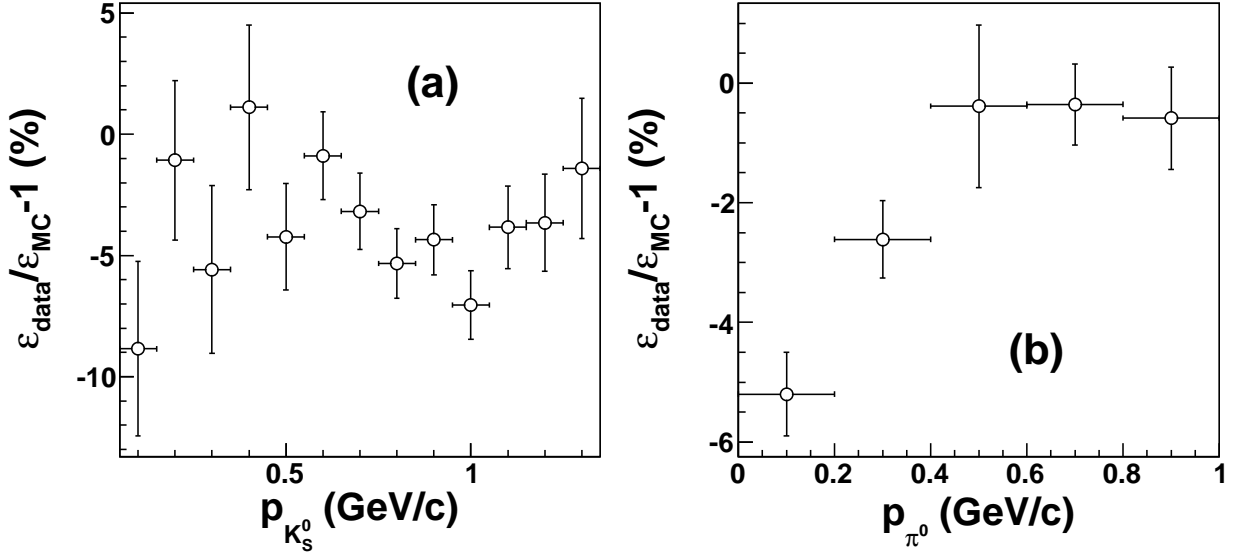}
  \caption{The differences of efficiencies between data and MC as a function of momentum, (a) for $K^0_S$ and (b) for $\pi^0$, for the control samples described in the text.}
  \label{fig:effdiff}
\end{figure}

A deviation of efficiency between data and MC simulation will cause a deviation of the fit results. Therefore, a momentum-dependent correction is applied to the final results. First, the differences of efficiencies between MC and data are determined. For the charged $\pi$ tracking efficiency and PID, Ref.~\cite{effdiff} has studied their momentum-dependent differences through $\psi^\prime\rightarrow \pi^+\pi^-J/\psi$ and $J/\psi\rightarrow \rho\pi\rightarrow \pi^+\pi^-\pi^0$. They are also studied using $D^0\rightarrow K^-\pi^+$ and $D^+ \rightarrow K^-\pi^+\pi^+$ control samples. The momentum-dependent differences in this range are all smaller than 2\% and are used to correct MC efficiencies. The $K^0_S$ efficiency is studied through $J/\psi\rightarrow K^{*-}K^+$ and $D^0\rightarrow K^{*-}\pi^+$ control samples. Besides the sample obtained by the standard selection, a loose selection without the $K^0_S$ requirement is used to obtain a reference sample. The distributions of missing mass squared of these $K^0_S$ are fitted with the shape of MC signal convolved by a Gaussian function plus the shape of the MC backgrounds. The number of expected events $N_{exp}$ is obtained from the reference sample, and the number of observed events $N_{obs}$ from the standard sample. Then the efficiency is taken as $N_{obs}/N_{exp}$. Dividing the samples into sub-samples according to momentum, momentum-dependent efficiencies are obtained. The same process is performed on data and MC events respectively, and their difference is shown in Fig.~\ref{fig:effdiff}(a). The $\pi^0$ efficiency is studied through the $D^0\rightarrow K\pi\pi^0$ control sample, and similar steps are taken. Figure \ref{fig:effdiff}(b) shows the difference in the $\pi^0$ reconstruction efficiency. According to the momentum-dependent differences, a correction is performed. Details of the correcting process are described in Appendix \ref{sec:correction}. The corrected results of the model D are listed in Table \ref{tab:isobarerr}.

In fits with these models,the $\overline{\kappa}$ is represented with a complex pole form, and the position of the pole $\kappa$ is allowed to float as a free complex parameter. The pole of the $\kappa$ is measured at $(752\pm15\pm69^{+55}_{-73},-229\pm21\pm44^{+40}_{-55})$~MeV, where the errors are statistical, experimental, and modeling uncertainties, respectively, consistent with the model C result of CLEO-c \cite{CLEOc}.

The mass and width of the $K^*_0(1430)^0$ are also floated, since the measured values from E791 \cite{E791a} and CLEO-c \cite{CLEOc} in the $D^+\rightarrow K^-\pi^+\pi^+$ decay are significantly different from
the measurement from the $Kp$ experiment LASS \cite{LASS}. In our fit, the mass and width of the $K^*_0(1430)^0$ are $1464\pm6\pm9^{+9}_{-28}$~MeV and $190\pm7\pm11^{+6}_{-26}$~MeV, respectively, consistent with the measurements from CLEO-c and E791. In our model without the $\overline{\kappa}$, the efficiency corrected results are $1444\pm4$~MeV and $283\pm11$~MeV, with statistical errors only.

\subsection{Cross-check with MIPWA}

The biggest issue of any Dalitz plot analysis is its model dependence. An attempt to mitigate the model dependence for the $D^+\rightarrow K^-\pi^+\pi^+$ decay under study is described in \cite{E791b}. Here, we apply this model-independent partial wave analysis (MIPWA) technique as a cross-check of our model D for the contributions of $K^{0}_{S}\pi^{0}$ $S$-wave.

The complex term $\mathcal{W}_{R}$ and $c_{NR}$ in Eq. \ref{eqn:waves} and \ref{eqn:res} can be used alone or in combination with other terms. In this check, it represents a correction to the complex amplitude of the isobar model. We use this term in the form of an $s$-dependent complex number
\begin{equation}
\mathcal{W}_{L,binned}(s)=a_L(s)e^{i\phi_L(s)},
\label{eqn:mipwa}
\end{equation}
with the functions $a_L(s)$ and $\phi_L(s)$ calculated by a linear interpolation between the bins for the magnitude $a_{Lk}$ and phase $\phi_{Lk}$, where $k(s)=1,2,...,N_L$ is an $s$-dependent index of these bins.

\begin{figure}
  \centering
  \includegraphics[width=1\linewidth]{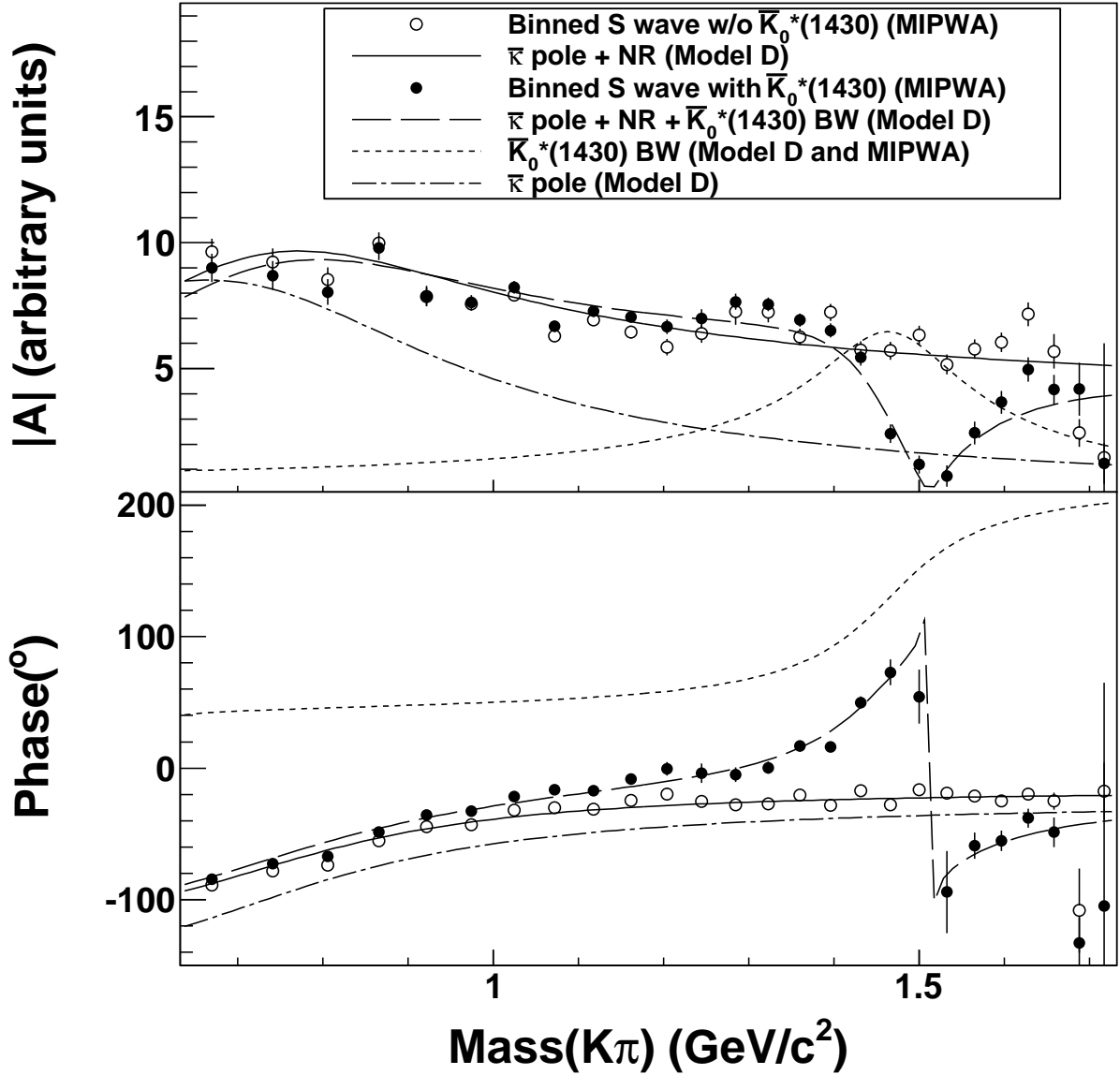}
  \caption{The magnitude and phase of the $K\pi$ $S$-wave in model D and the MIPWA. The open circles with error bars (statistical uncertainties only) show the binned $K\pi$ $S$-wave without the $K^*_0(1430)$ and the black dots show the total $K\pi$ $S$-wave. Other curves show the $S$-wave components of model D.}
  \label{fig:mipwa}
\end{figure}

We test two models, one with a binned $K^0_S\pi^0$ $S$-wave, and another with a binned $K^0_S\pi^0$ $S$-wave excluding the $\overline{K}^*_0(1430)^0$ (whose contribution is kept in its Breit-Wigner form). The measured $S$-wave magnitudes and phases are illustrated in Fig.~\ref{fig:mipwa}. In order to compare with the previous $D^+ \rightarrow K^-\pi^+\pi^+$ results, we measure all magnitudes and phases relative to the $\overline{K}^*(892)^0\pi^+$ decay mode in the MIPWA fits. Comparing the binned $S$-wave fit without the $\overline{K}^*_0(1430)^0$ component to a sum of the $\overline{\kappa}$ pole and the non-resonant component in the model D, and the total binned $S$-wave to a sum of the $\overline{\kappa}^0$ pole, $\overline{K}^*_0(1430)^0$ and the non-resonant component in the model D, respectively, these models are consistent with the model-dependent analysis. It is obvious that there is still a phase variation from low mass threshold to higher mass in the $K^0_S\pi^0$ $S$-wave excluding the $\overline{K}^*_0(1430)^0$, similar with the combination of the $NR$ and the $\overline{\kappa}\pi^+$ in model D. In the total binned $K^0_S\pi^0$ $S$-wave, the amplitude is distorted by a contribution from the $\overline{K}^*_0(1430)$ resonance.

\section{Systematic Uncertainties}
\label{syserr}

\begin{table*}
  \centering
  \tablesize
  \caption{A summary of the statistical and systematic errors on the fit parameters of the model D. The \textquotedblleft Value\textquotedblright\,and \textquotedblleft Statistical\textquotedblright\,columns show the results from the momentum-dependent efficiency correction. The three columns under \textquotedblleft Experimental Errors\textquotedblright\, (\textquotedblleft Modeling Errors\textquotedblright) summarize the systematic uncertainties due to experimental (modeling) sources respectively, described in the text in detail. The $S$-wave is calculated by adding the non-resonant component, the $\overline{\kappa}^0\pi^+$, and the $\overline{K}^*_0(1430)^0\pi^+$.\label{tab:isobarerr}}
  \begin{tabularx}{\linewidth}{|c|X|X|XXX|XXX|}
  \hline
  \hline
  Parameters & Value & Statistical & \multicolumn{3}{c|}{Experimental Errors} & \multicolumn{3}{c|}{Modeling Errors} \\
  \cline{4-9}
    &   & Errors & Background & Efficiency & Total & Shape & Add & Total \\
  \hline
  $NR$ FF(\%)                         & 4.6   & 0.7 & 3.5  & 1.0  & 3.6  & $_{-1.5}^{+2.9}$   & $_{-3.3}^{+2.7}$   & $_{-3.6}^{+4.0}$ \\
  $NR$ Phase(\ensuremath{^{\circ}})                           & 279 & 6 & 5  & 15 & 15 & $_{-25}^{+6}$  & $_{-12}^{+22}$ & $_{-27}^{+23}$  \\
  $\rho(770)^+$ FF(\%)              & 83.4  & 2.2 & 2.7  & 0.7  & 2.8  & $_{-1.9}^{+1.1}$   & $_{-1.1}^{+6.4}$   & $_{-2.2}^{+6.5}$  \\
  $\rho(1450)^+$ FF(\%)             & 2.1   & 0.3 & 0.9  & 0.9  & 1.2  & $_{-0.1}^{+0.7}$   & $_{-1.5}^{+0.8}$   & $_{-1.5}^{+1.0}$  \\
  $\rho(1450)^+$ Phase(\ensuremath{^{\circ}})               & 187 & 3 & 4  & 4  & 5  & $_{-15}^{+9}$  & $_{-5}^{+26}$  & $_{-16}^{+28}$  \\
  $\overline{K}^*(892)^0$ FF(\%)    & 3.58   & 0.17 & 0.12  & 0.11  & 0.17  & $_{-0.18}^{+0.31}$   & $_{-0.28}^{+0.16}$   & $_{-0.34}^{+0.35}$  \\
  $\overline{K}^*(892)^0$ Phase(\ensuremath{^{\circ}})      & 293 & 2 & 1  & 2  & 2  & $_{-7}^{+2}$   & $_{-2}^{+6}$   & $_{-7}^{+6}$  \\
  $\overline{K}^*_0(1430)^0$ FF(\%) & 3.7   & 0.6 & 0.6  & 0.5  & 0.8  & $_{-0.3}^{+0.4}$   & $_{-0.8}^{+0.7}$   & $0.8$  \\
  $\overline{K}^*_0(1430)^0$ Phase(\ensuremath{^{\circ}})   & 334 & 5 & 8  & 4  & 9  & $_{-10}^{+1}$   & $_{-28}^{+3}$  & $_{-30}^{+3}$  \\
  $\overline{K}^*(1680)^0$ FF(\%)   & 1.3   & 0.2 & 0.6  & 0.2  & 0.7  & $_{-0.1}^{+0.6}$   & $_{-1.1}^{+0.1}$   & $_{-1.1}^{+0.6}$  \\
  $\overline{K}^*(1680)^0$ Phase(\ensuremath{^{\circ}})     & 252 & 2 & 9  & 6  & 11 & $_{-2}^{+6}$   & $_{-28}^{+7}$  & $_{-28}^{+9}$  \\
  $\overline{\kappa}^0$ FF(\%)                              & 7.7   & 1.2 & 2.5  & 3.1  & 4.0  & $_{-2.7}^{+2.0}$   & $_{-0.1}^{+4.7}$   & $_{-2.7}^{+5.1}$  \\
  $\overline{\kappa}^0$ Phase(\ensuremath{^{\circ}})        & 93  & 7 & 25 & 14 & 28 & $_{-7}^{+14}$  & $_{-22}^{+16}$ & $_{-23}^{+21}$  \\
  \hline
  $NR$+$\overline{\kappa}^0$ FF(\%)         & 18.6  & 1.7 & 1.1  & 1.0  & 1.5  & $_{-3.7}^{+1.6}$   & $_{-2.3}^{+0.5}$   & $_{-4.4}^{+1.7}$           \\
  $K^0_S\pi^0$ $S$-wave FF(\%)            & 17.3  & 1.4 & 2.1  & 0.5  & 2.1  & $_{-3.8}^{+0.7}$   & $_{-0.6}^{+2.6}$   & $_{-3.8}^{+2.7}$           \\
  \hline
  \hline
  \end{tabularx}
\end{table*}

In our analysis, according to Eq. \ref{eqn:probability}, there are several possible sources of systematic uncertainties: the background, the efficiency, the numerical integration, and the modeling of the decay. In order to estimate systematic uncertainties of the fit parameters due to these sources, we carry out the checks described in this section in detail. We require $10^{-8}$ precision to get the integral of the p.d.f. If we improve the precision by an order of magnitude, we find negligible change. The final systematic errors are shown in Table \ref{tab:isobarerr}. The \textquotedblleft Total\textquotedblright\,experimental errors are obtained as a quadratic sum of that from background and efficiency.

\subsection{Background}
\label{bkgerr}
The uncertainties from the background (shown in the \textquotedblleft Background\textquotedblright\,column) come from two sources: the background shape and the background normalization. The background shape depends on both the parameterization and the sideband approximation.

There is a difference between the true background shape and the polynomial function as pointed out in Section \ref{background}. But in the high-mass sideband, we do not know the shape of the background component because of the signal tail. According to Fig.~\ref{fig:bkgfit}, the differences are close in cases of low-mass sideband, high-mass sideband, and combined sideband. Hence we choose the low-mass sideband to examine the 3rd order polynomial parameterization. Inputting the low-mass sideband shape using a histogram p.d.f., we compare to the fit result with the parameterized low-mass sideband shape. We take the variation as the systematic error.

Both sidebands are used to parameterize background in the final fit, and it is believable that the deviation of this shape from the real background would not exceed the difference between backgrounds in the low-mass and high-mass sidebands. Inputting the background shape parameterized by these two sidebands, the difference of results is estimated as the uncertainty due to the background shape.

In Section \ref{sec:data}, we estimate that the statistical error of the signal ratio is $0.1\%$. Through comparing MC truth to the result of fitting on the $m_{rec}$ distribution of MC sample, its systematic uncertainty is estimated to be $^{+0.1}_{-1.4}\%$, and if the signal ratio is floated in the Dalitz fit, the fitted value is $(83.3\pm0.4)\%$. They are consistent with each other. We change the signal ratio in the fit to change the background level by one standard deviation. The variation of results is taken as the estimation of uncertainty of the background level.

\subsection{Efficiency}
The systematic uncertainty from the efficiency (shown in the \textquotedblleft Efficiency\textquotedblright\,column) includes two terms: the efficiency parameterization and the difference between data and MC. The sources of the difference of data and MC include event selection criteria, tracking, unstable particle reconstruction, and particle identification. The resolution of the detector is also considered here.

For the efficiency parameterization, we change the global polynomial fit to the average efficiencies of local bins. Each bin's efficiency value is replaced by the average efficiency. We also try smoothing the efficiencies by averaging either nine or twenty-five nearest neighbors as a check. The differences caused by using different parameterizations is also considered in the systematic error.

Another efficiency parameterization, which is obtained using a MC sample uniform in phase space, is used as a cross check. The variation is taken as one of the systematic uncertainties.

Because the resolutions of $\Delta E$ and $m_{rec}$ in data are a little larger than in MC, the efficiency shape could possibly be different as well. In order to estimate the uncertainty caused by the cuts, we change the cuts on the MC sample to make the cumulative probability at the cut position the same as data. This check indicates that this uncertainty is small.

The particle reconstruction and identification are also possible sources of systematic error. If the differences between data and MC are independent of 3-momentum, there will be no effect on the relative branching fractions. Therefore, a momentum dependent correction on reconstruction and PID efficiency is performed, as described in Appendix \ref{sec:correction}. Correspondingly, the r.m.s. of the measured values are taken as an estimate of the systematic errors.

\begin{table*}
  \centering
  \caption{Partial branching fractions calculated by combining our fit fractions with the PDG's $D^+\rightarrow K^0_S\pi^+\pi^0$ branching ratio. The errors shown are statistical, experimental systematic, and modeling systematic, respectively.}\label{tab:partial}
  \begin{tabularx}{\linewidth}{|X|X|}
  \hline
  \hline
  Mode & Partial Branching Fraction (\%) \\
  \hline
  $D^+\rightarrow K^0_S\pi^+\pi^0$ Non Resonant & 0.32$\pm$0.05$\pm$0.25$^{+0.28}_{-0.25}$ \\
  $D^+\rightarrow \rho^+ K^0_S,\rho^+\rightarrow \pi^+\pi^0$  & 5.83$\pm$0.16$\pm$0.30$^{+0.45}_{-0.15}$ \\
  $D^+\rightarrow \rho(1450)^+ K^0_S,\rho(1450)^+ \rightarrow \pi^+\pi^0$  & 0.15$\pm$0.02$\pm$0.09$^{+0.07}_{-0.11}$ \\
  $D^+\rightarrow \overline{K}^*(892)^0\pi^+,\overline{K}^*(892)^0\rightarrow K^0_S\pi^0$ & $0.250\pm0.012\pm0.015^{+0.025}_{-0.024}$\\
  $D^+\rightarrow \overline{K}^*_0(1430)^0\pi^+,\overline{K}^*_0(1430)^0\rightarrow K^0_S\pi^0$& $0.26\pm0.04\pm0.05\pm0.06$\\
  $D^+\rightarrow \overline{K}^*(1680)^0\pi^+,\overline{K}^*(1680)^0\rightarrow K^0_S\pi^0$& $0.09\pm0.01\pm0.05^{+0.04}_{-0.08}$\\
  $D^+\rightarrow \overline{\kappa}^0\pi^+,\overline{\kappa}^0\rightarrow K^0_S\pi^0$& $0.54\pm0.09\pm0.28^{+0.36}_{-0.19}$\\
  \hline
  $NR$+$\overline{\kappa}^0\pi^+$&1.30$\pm$0.12$\pm$0.12$^{+0.12}_{-0.30}$\\
  $K^0_S\pi^0$ $S$-wave & 1.21$\pm$0.10$\pm$0.16$^{+0.19}_{-0.27}$\\
  \hline
  \hline
\end{tabularx}
\end{table*}

To estimate the experimental systematic error due to the finite resolution of the Dalitz plot variables, we have included the effects of smearing when fitting the data as a check. This was done by measuring the resolution as a function of position across the Dalitz plot and numerically convoluting this with the amplitude at each point when performing the fit. The resulting change of parameters from the nominal best fit is very small and can be neglected when compared to other uncertainties.

\subsection{Model}
Systematic uncertainties of the modeling of the decay can arise from the parameterization of the resonances (shown in the \textquotedblleft Shape\textquotedblright\,column), which include barrier factors, dynamical functions and resonance parameters, and also come from the choice of resonances in the baseline fit (shown in the \textquotedblleft Add\textquotedblright\,column). The \textquotedblleft Shape\textquotedblright\,and \textquotedblleft Add\textquotedblright\,columns are added in quadrature to obtain the final model dependent systematic errors, shown in the \textquotedblleft Total\textquotedblright\,column under \textquotedblleft Modeling Errors\textquotedblright.

We test the exponential barrier factor $F^0_V=e^{-(q^2-q^2_V)/12}$ as an alternative description of the scalar intermediate resonances in Table \ref{tab:form}. A smaller $NR$ fraction is obtained, but the total $K\pi$ $S$-wave is relatively unaffected. We do not consider it as a systematic error. We also test the fit by changing the radial parameters used in the barrier factors from 0~GeV$^{-1}$ to 3~GeV$^{-1}$ for the intermediate resonances, and from 0~GeV$^{-1}$ to 10~GeV$^{-1}$ for the $D^+$ meson. The maximum likelihood values $\mathcal{L}$ appear at $r_D\approx 2.75\text{~GeV}^{-1}$ and $r_R\approx 1.48\text{~GeV}^{-1}$, respectively. It indicates that the radial parameter of the intermediate resonances is consistent with 1.5~GeV$^{-1}$ and the radius of the $D$ meson has large uncertainty. The variation caused by the uncertainties of the radii is taken as a systematic error.

Different resonance shapes for $\overline{K}^*_0(1430)$ and $\overline{\kappa}$ are tested. A Flatt\'{e} form for $\overline{K}^*_0(1430)$ and Breit-Wigner for $\overline{\kappa}$ are tried. If only the $\overline{K}^*_0(1430)$ is changed to the Flatt\'{e} form, the $\chi^2$ changes by $-11$ units. If only the $\overline{\kappa}$ is changed to Breit-Wigner, the $\chi^2$ changes by 11. We also perform the fit while floating the masses and widths of the $\rho(770)$ and the $\overline{K}^*(892)$. The variations from the nominal values are taken as an estimation of this systematic uncertainty.

The final systematic check is on our choice of which resonances are to be included. We do two fits for different $\rho(1450)^+$ and $\rho(1700)^+$, and take the variation of parameters as the error. We also add insignificant resonances one by one, including $\overline{K}^*(892)^+\pi^0$, $\overline{K}^*(1410)^0\pi^+$, $\overline{K}^*_0(1430)^+\pi^0$, $\overline{K}^*_2(1430)^0\pi^+$, $\overline{K}^*_3(1780)^0\pi^+$, and watch the variations of the fit fractions of the observed channels, which is taken as an additional systematic uncertainty.

\section{Summary And Conclusions}
\label{summary}
We describe an amplitude analysis of the $D^+ \rightarrow K^0_S\pi^+\pi^0$ Dalitz plot. We start with a BESIII data set of 2.92~fb$^{-1}$ of \epem collisions accumulated at the peak of the $\psi(3770)$, and select 166694 candidate events with a background of $(15.1\pm0.1^{+1.4}_{-0.1})\%$.

We fit the distribution of data to a coherent sum of six intermediate resonances plus a non-resonant component, with a low mass scalar resonance, the $\overline{\kappa}$, included. The final fit fraction and phase for each component is given in Table \ref{tab:isobarerr}. These fit fractions, multiplied by the world average $D^+ \rightarrow K^0_S\pi^+\pi^0$ branching ratio of (6.99$\pm$0.27)\% \cite{PDG}, yield the partial branching fractions shown in Table \ref{tab:partial}. The error on the world average branching ratio is incorporated by adding it in quadrature with the experimental systematic errors on the fit fractions to give the experimental systematic error on the partial branching fractions.

In this result, the $K^0_S\pi^0$ waves can be compared with the $K^-\pi^+$ waves in the $D^+\rightarrow K^-\pi^+\pi^+$ decay. For example, according to our measured branching ratio of $D^+\rightarrow \overline{K}^{*0}\pi^+\rightarrow K^0_S\pi^+\pi^0$ and the PDG value of branching ratio of $D^+\rightarrow \overline{K}^{*0}\pi^+\rightarrow K^-\pi^+\pi^+$ of (1.01$\pm$0.11)\%, the ratio of the branching fractions of $D^+\rightarrow\overline{K}^{*0}\pi^+\rightarrow K^-\pi^+\pi^+$ and $D^+\rightarrow\overline{K}^{*0}\pi^+\rightarrow\overline{K}^0\pi^+\pi^0$ is calculated to be $2.02\pm0.34$, which is consistent with the expectation.

We also apply a model-independent approach to describe the Dalitz plot, developed in Ref. \cite{E791b}, to confirm the results. The $K\pi$ $S$-wave can be well-described by a $\overline{\kappa}$, a $\overline{K}^*_0(1430)$, and a non-resonant component. The resonance parameters of the $\kappa$ and the $K^*_0(1430)$ are consistent with the results of E791 \cite{E791a} and CLEO-c \cite{CLEOc} in the $D^+\rightarrow K^-\pi^+\pi^+$ decay.

\begin{acknowledgments}
The BESIII collaboration thanks the staff of BEPCII and the computing center for their strong support. This work is supported in part by the Ministry of Science and Technology of China under Contract No. 2009CB825200; Joint Funds of the National Natural Science Foundation of China under Contracts Nos. 11079008, 11179007, U1332201; National Natural Science Foundation of China (NSFC) under Contracts Nos. 10625524, 10821063, 10825524, 10835001, 10935007, 11125525, 11235011; the Chinese Academy of Sciences (CAS) Large-Scale Scientific Facility Program; CAS under Contracts Nos. KJCX2-YW-N29, KJCX2-YW-N45; 100 Talents Program of CAS; German Research Foundation DFG under Contract No. Collaborative Research Center CRC-1044; Istituto Nazionale di Fisica Nucleare, Italy; Ministry of Development of Turkey under Contract No. DPT2006K-120470; U. S. Department of Energy under Contracts Nos. DE-FG02-04ER41291, DE-FG02-05ER41374, DE-FG02-94ER40823, DESC0010118; U.S. National Science Foundation; University of Groningen (RuG) and the Helmholtzzentrum fuer Schwerionenforschung GmbH (GSI), Darmstadt; WCU Program of National Research Foundation of Korea under Contract No. R32-2008-000-10155-0. This paper is also supported by the NSFC under Contract Nos. 10875138, 11205178.
\end{acknowledgments}

\appendix
\section{Isobar Model}
\label{sec:isobar}
In general, for the decay of $D\rightarrow Rc$, $R\rightarrow ab$, where the spins of $a$, $b$, and $c$ are equal to zero, the orbital angular momentum between $R$ and $c$ is equal to the spin of $R$ and the angular distribution can be simplified to a function of the momentum of $a$ ($p_a$) and the momentum of $c$ ($p_c$) in the $R$ rest frame:
\begin{equation}
Z_L=(-2p_a p_c)^LP_L(cos\theta),
\end{equation}
where the Legendre polynomials $P_L(\cos\theta)$ depend on the orbital angular momentum (the spin of R). Here $\theta$ is the helicity angle and its cosine is given in terms of the masses $m_a$($m_c$) and energies $E_a$($E_c$) of the $a$($c$) in the $R$ rest frame:
\begin{equation}
\cos\theta=\frac{m_a^2+m_c^2+2E_a E_c-m_{ac}^2}{2p_a p_c}.
\end{equation}

In this analysis, intermediate resonances are parameterized using the standard Breit-Wigner function defined as
\begin{equation}
W_R\left(m_{ab}\right)=\frac{1}{m^2_R-m^2_{ab}-im_{R}\Gamma(m_{ab})},
\end{equation}
where $m_R$ is the resonance mass and $m_{ab}$ is the invariant mass of the $ab$ system, and the mass-dependent width $\Gamma(m_{ab})$ has the usual form \cite{width}:
\begin{equation}
\label{eqn:gamma}
\Gamma(m_{ab})=\Gamma_R\left(\frac{p_a}{p_R}\right)^{2L+1}\left(\frac{m_R}{m_{ab}}\right)\left(F^L_R\right)^2,
\end{equation}
where $\Gamma_R$ is the resonance width, and $p_R$ is the value of $p_a$ when $m_{ab}=m_R$.

In Eq. \ref{eqn:element}, Eq. \ref{eqn:res}, and Eq. \ref{eqn:gamma}, $F^L_D$ and $F^L_R$ are the barrier factors for the production of $Rc$ and $ab$, defined using the Blatt-Weisskopf form \cite{Blatt-Weisskopf}, as listed in Table \ref{tab:form}.

\begin{table}
  \centering
  \caption{The Blatt-Weisskopf barrier factor used in this analysis. The index V stands for the $D$ or R decay vertex, and $q=r_Vp$ ($p$ is the magnitude of the momentum of the decay daughters in the rest frame of mother particle, and $r_V$ is the effective radius for the $D$ or R vertex). For both $D$ and R decays, $q_V=r_Vp_V$, where $p_V$ is the value of $p$ when $m_{ab}=m_R$.}\label{tab:form}
  \begin{tabularx}{0.48\textwidth}{XX}
    \hline
    L & Form factor $F^L_V$                     \\
    \hline
    0 & 1                                     \\
    1 & $\sqrt{\frac{1+q^2_V}{1+q^2}}$                          \\
    2 & $\sqrt{\frac{9+3q^2_V+q^4_V}{9+3q^2+q^4}}$                 \\
    3 & $\sqrt{\frac{405+45q^2_V+6q^4_V+q^6_V}{405+45q^2+6q^4+q^6}}$       \\
    \hline
  \end{tabularx}
\end{table}

For the $\overline{\kappa}$ we have tested both the Breit-Wigner function and the complex pole proposed in Ref. \cite{kappa}:
\begin{equation}
W_R(m_{ab})=\frac{1}{s_R-m^2_{ab}}=\frac{1}{m^2_R-m^2_{ab}-im_R\Gamma_R},
\end{equation}
which is equivalent to a Breit-Wigner function with constant width. In the fit,
\begin{equation}
\label{eqn:kappa}
s_R=(\Re+i \Im)^2,
\end{equation}
where $\Re$ and $\Im$ are the two parameters of the complex pole.

\section{Momentum-Dependent Correction}
\label{sec:correction}
Based on momentum-dependent differences in efficiency, we can correct the MC efficiency to the expected data efficiency through a sampling method, and use the corrected efficiency to improve the results. The detailed steps are described as follows. First, an MC sample is generated and selected using the same event selection as data, and its events are denoted as $E_i(p_{K^0_S},p_{\pi^+},p_{\pi^0}),i=1\ldots N$, where $p_{K^0_S}$, $p_{\pi^+}$, and $p_{\pi^0}$ are the momentum of $K^0_S$, $\pi^+$, and $\pi^0$, respectively. Before sampling, the efficiency ratio of data and MC is computed as
\begin{equation}
r_{\varepsilon}(p_{K^0_S},p_{\pi^+},p_{\pi^0})=\prod_c \frac{\varepsilon_{c,data}(p_{c})}{\varepsilon_{c,MC}(p_{c})},
\end{equation}
where the subscripts $c$ include the $K^0_S$ efficiency, the $\pi^0$ efficiency, the $\pi^+$ tracking efficiency, and the $\pi^+$ PID efficiency, and $p_{c}$ denotes the momentum of the corresponding particles. Then for each event $E_i$, if $r_{\varepsilon}$ is less than one, it will be compared with a uniform (0,1) random number $\zeta$, and the event is kept only if $r_{\varepsilon}>\zeta$; if $r_{\varepsilon}$ is larger than one, the event will always be kept, and it will be repeated once while $r_{\varepsilon}-1>\zeta$. The sampling process is complete after all selected events are looped over. Finally, the efficiency parameterization is applied to the sampled events and the new efficiency parameters are used to fit data.

To remove the statistical fluctuations while sampling and the uncertainty of measurement of $r_{\varepsilon}$, we repeat this process, and change the $r_{\varepsilon}$ value according to its uncertainty each time. Then we can obtain the distribution of results following a Gaussian distribution. The means denote the corrected results, and the sigmas describe the uncertainty of sampling and the measurement of efficiencies.

\end{document}